\documentclass[reprint,amsmath,amssymb,aps,superscriptaddress]{revtex4-1}
\usepackage{hyperref}
\usepackage[hyphenbreaks]{breakurl}

\newcommand{\myemail}{pablo.marchant@kuleuven.be }
\usepackage{amsmath}
\usepackage{dcolumn}
\usepackage{bm}
\usepackage{graphicx}
\usepackage{color}
\usepackage[normalem]{ulem}
\newcommand*\aap{A\&A}

\newcommand*\aj{AJ}
\newcommand*\apjl{ApJ}

\newcommand*\apjs{ApJS}

\newcommand*\apss{Ap\&SS}
\newcommand*\araa{ARA\&A}

\newcommand*\mnras{MNRAS}

\newcommand*\pasa{PASA}

\begin{document}


\title{Eclipses of continuous gravitational waves as a probe of stellar structure}


\author{Pablo Marchant}
\email{\myemail}
\affiliation{Center for Interdisciplinary Exploration and Research in
Astrophysics (CIERA), Northwestern University, 1800 Sherman Avenue, Evanston, IL 60208, USA;\\
Department of Physics and Astronomy, Northwestern University, 2145 Sheridan
Road, Evanston, IL 60208, USA}
\affiliation{Institute of Astrophysics, KU Leuven, Celestijnenlaan 200 D, 3001 Leuven, Belgium}
\author{Katelyn Breivik}
\affiliation{Canadian Institute for Theoretical Astrophysics, University of
Toronto, 60 St. George Street, Toronto, Ontario, M5S 1A7, Canada}
\affiliation{Center for Interdisciplinary Exploration and Research in
Astrophysics (CIERA)
and Department of Physics and Astronomy, Northwestern University, 2145 Sheridan
Road, Evanston, IL 60208, USA}
\author{Christopher P L Berry}
\affiliation{Center for Interdisciplinary Exploration and Research in
Astrophysics (CIERA), Northwestern University, 1800 Sherman Avenue, Evanston, IL 60208, USA;\\
Department of Physics and Astronomy, Northwestern University, 2145 Sheridan
Road, Evanston, IL 60208, USA}
\author{Ilya Mandel}
\affiliation{Monash Centre for Astrophysics, School of Physics and Astronomy,
Monash University, Clayton, Victoria 3168, Australia;\\ 
ARC Centre of Excellence for Gravitational Wave Discovery -- OzGrav}
\affiliation{Birmingham Institute for Gravitational Wave Astronomy and School of
Physics and Astronomy, University of Birmingham, Birmingham, B15 2TT, United
Kingdom}
\author{Shane L. Larson}
\affiliation{Center for Interdisciplinary Exploration and Research in
Astrophysics (CIERA), Northwestern University, 1800 Sherman Avenue, Evanston, IL 60208, USA;\\
Department of Physics and Astronomy, Northwestern University, 2145 Sheridan
Road, Evanston, IL 60208, USA}

\date{\today}

%
%




\begin{abstract}
   Although gravitational waves only interact weakly with matter, their
   propagation is affected by a gravitational potential. If a gravitational wave source is
   eclipsed by a star, measuring these perturbations provides a way to directly
   measure the distribution of mass throughout the stellar interior. We compute
   the expected Shapiro time delay, amplification and deflection during an
   eclipse, and show how this can be used to infer the mass distribution of the
   eclipsing body. We identify continuous gravitational waves from neutron stars as the best
   candidates to detect this effect. When the Sun eclipses a far-away source,
   depending on the depth of the eclipse the time-delay can change by up to
   $\sim 0.034~\mathrm{ms}$, the gravitational-wave strain amplitude can increase by $\sim 4\%$,
   and the apparent position of the source in the sky can vary by $4''$.
   Accreting neutron stars with Roche-lobe filling companion stars have a high
   probability of exhibiting eclipses, producing similar time delays but
   undetectable changes in amplitude and sky location. Even for the most rapidly
   rotating neutron stars, this time delay only corresponds to a few percent of
   the phase of the gravitational wave, making it an extremely challenging measurement. However,
   if sources of continuous gravitational waves exist just below the limit of detection of
   current observatories, next-generation instruments will be able to observe
   them with enough precision to measure the signal of an eclipsing star.
   Detecting this effect would provide a new direct probe to the interior of
   stars, complementing asteroseismology and the detection of solar neutrinos.
\end{abstract}

\maketitle

\section{Introduction}

The subject of lensing of gravitational waves (GWs) was studied in
the early 1970s and 1980s in the context of amplifying possible signals to the
point of detection. This was in part driven by claims 
of the observation of GWs using cylindrical bar detectors \cite{Weber1969}, for which the
reported amplitude was too high to be explained by astrophysical sources.
Considering the Galactic core as a lens, it was shown that this
was insufficient to explain those detections \cite{Lawrence1971, Ohanian1973}. Lensing by the
Sun was also shown to be unimportant for the observation of GWs, as diffraction effects imply
that a significant amplification of the signal is only
expected for GWs with frequencies $>10^4~\mathrm{Hz}$ \cite{Ohanian1973,BontzHaugan1981,
Ohanian1983}. This is higher than the frequencies of known astrophysical GW sources, which are
not expected to exceed a few kilohertz \cite{Kokkotas2008}.
Currently, strong lensing is only expected to affect a
small number of observable sources of GWs \cite{Smith+2017,Ng+2018,Shun-Sheng+2018}, 
and there is no strong evidence for current detections having been strongly
lensed \cite{Hannuksela+2019}. Microlensing of GWs is also considered to be an
unlikely event, but owing to the relatively low frequencies of GW sources can
lead to wave-optical phenomena that allow the inference of additional
information about the lens \cite{TakahashiNakamura2003,Moylan+2008,Liao+2019}.

Even if GW lensing is not expected to play a role in the majority of observable
sources, measuring small effects of intervening matter on GWs can provide
interesting astrophysical information.  The detection of GWs from merging binary
black holes (BHs) \cite{Abbott_GW150914_2016,Abbott+2019_GWTC-1} and neutron
stars (NSs) \cite{Abbott+2017_GW170817,Abbott+2019_GWTC-1} by the Advanced LIGO
\cite{Aasi+2015} and Virgo \cite{Acernese+2015} detectors
makes it possible to use them as astrophysical tools.
In particular, GWs crossing the interior
of a star carry information on its internal mass distribution. For instance, it has been
proposed that measuring the deflection angle of a GW source eclipsed by the Sun 
will yield the solar density profile \cite{CyranskiLubkin1974}.

In this paper
we discuss the effects of eclipsing stars on GWs, and how these provide
information on the interior of the eclipsing star. {Our focus is on
high-frequency ($>1$ Hz) GWs that are
potentially detectable by ground-based observatories.} In Section~\ref{c_vs_cbc} we discuss
different sources of GWs that could be used for this purpose,
and show that high-frequency continuous GWs (CWs) work best.
In Section~\ref{effects} we analyze the effects produced on high-frequency GWs
crossing the interior of the Sun using both geometric and wave
optics, while in Section
\ref{sec:eclipse_binary} we discuss the case where the eclipsing star is a
binary companion to the source of GWs. We explore the detectability of
these effects in Section~\ref{sec:detect}, and give our conclusions in Section
\ref{conclusions}.  All code used to produce figures and compute our results is
available at
\footnote{\href{https://doi.org/10.5281/zenodo.2653899}{doi.org/10.5281/zenodo.2653899}}.

\section{Continuous GWs or compact binary coalescences}
\label{c_vs_cbc}
Although only GWs from compact binary coalescences
(CBCs) have been directly detected to date \cite{Abbott+2019_GWTC-1},
these sources are not useful for extracting information from an eclipsing event.
If we are interested in observing a source
behind the Sun, the probability of observing this for a CBC (assuming they are
isotropically distributed) is  given by
the fraction of the sky that is covered by the solar disk which has an
angular diameter of $\sim 32'$. This is because such sources pass quickly through
the ground-based detector band, and during this time the position of the Sun is essentially
static. The probability of it being located behind the Sun is then just
$0.00054\%$, and even after $10^4$ observations, there's only a $5.3\%$ chance
that at least one source is eclipsed.
\begin{figure}
   \begin{center}
      \includegraphics[width=1\columnwidth]{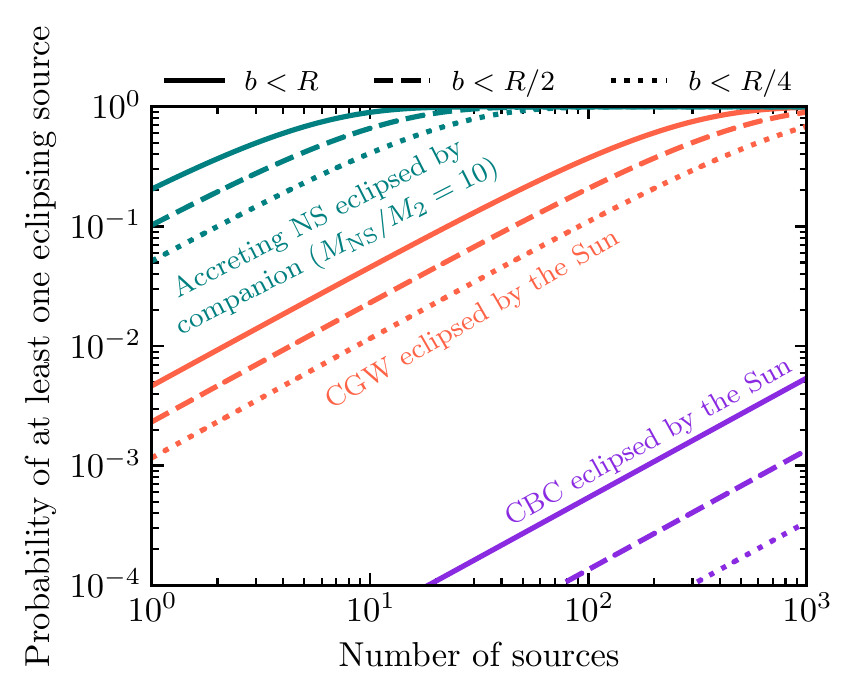}
      \caption{Probability of observing at least one eclipsing source as a
      function of the number of observed sources. Colors indicate different types of
      sources, while line styles indicate the depth of the eclipse. For the case
      of CBCs and CWs eclipsed by the Sun, probabilities are computed assuming an
      isotropic distribution in the sky, and an observer on Earth. Probabilities
      for the case of an accreting NS being eclipsed by its companion are
      computed assuming a mass ratio $q\equiv M_\mathrm{NS}/M_2=10$ and by approximating
      the stellar companion as a sphere with a radius equal to its Roche lobe.}
      \label{fig:numsources}.
   \end{center}
\end{figure}
Even if an eclipsing CBC is detected to high precision,  it is difficult to
distinguish effects inherent to the source from those produced by the eclipsing
star in the absence of previous information about the source. GW signals from a
CBC source are short-lived in the LIGO--Virgo band (a binary neutron star evolves
form a GW frequency of 10 Hz to merger in less than 20 minutes) relative to the
duration of an eclipse; this makes it implausible to compare the signal  of an
eclipsed source against its pre- or post-eclipse signal.

On the other hand, CWs are ideal for this purpose as any
source located $\lesssim 16'$ from the ecliptic and lasting more than a year will be
eclipsed by the Sun. The probability of a single source undergoing an annual
eclipse (assuming an isotropic distribution) is $0.47\%$, and with $200$ sources the
probability that at least one will undergo an eclipse is $61\%$ (see Fig.~\ref{fig:numsources}). The likelihood
of this happening is actually larger; unlike CBCs, many expected sources of
CWs are Galactic, thus not isotropically distributed in the sky,
and the Sun crosses the Galactic bulge. Moreover, a CW can be
studied and characterized \emph{before} a lensing event, making it easier to extract
information from an eclipse.

In terms of expected sources of CWs, binary white dwarfs will
be a prime source for the LISA observatory \cite{Amaro-Seoane+2017}, with known sources predicted to be
detectable given the design sensitivity of the instrument
\cite{StroeerVecchio2006,Kupfer+2018}.
However, the wavelengths of these sources are larger than the Sun, as are the arms
of the LISA constellation itself. Any effects produced
by the Sun on such long-wavelength sources are small due to
diffraction. As we show later, signals with GW
frequencies below $10~\mathrm{Hz}$ are essentially unperturbed by the Sun.

Rotating NSs are potential sources of high-frequency CWs
\cite{Prix2009,Riles2017}; a NS with a non-zero quadrupolar moment
is expected to emit GWs at a frequency $f_{\rm
GW}=2\nu$, where $\nu$ is the rotational frequency of the NS
\cite{ZimmermannSzedenits1979}. For known pulsars with measured time
derivatives, a rough upper limit on the strength of emitted GWs can be obtained
by assuming its spin-down is solely due to energy emitted in GWs. 
The latest searches for isolated sources using data from the first and second
observing runs of Advanced LIGO, have not resulted in a detection
\cite{Abbott+2019known_pulsars,Abbott+2019narrow_band,Abbott+2019blind}.
However, the searches made for known pulsars
\cite{Abbott+2019known_pulsars,Abbott+2019narrow_band} have further increased
the sample of young pulsars for which the spin-down limit is reached to $20$,
and are within factors of a few of the spin-down limit for some millisecond
pulsars (MSPs).

\begin{figure}
   \begin{center}
      \includegraphics[width=1\columnwidth]{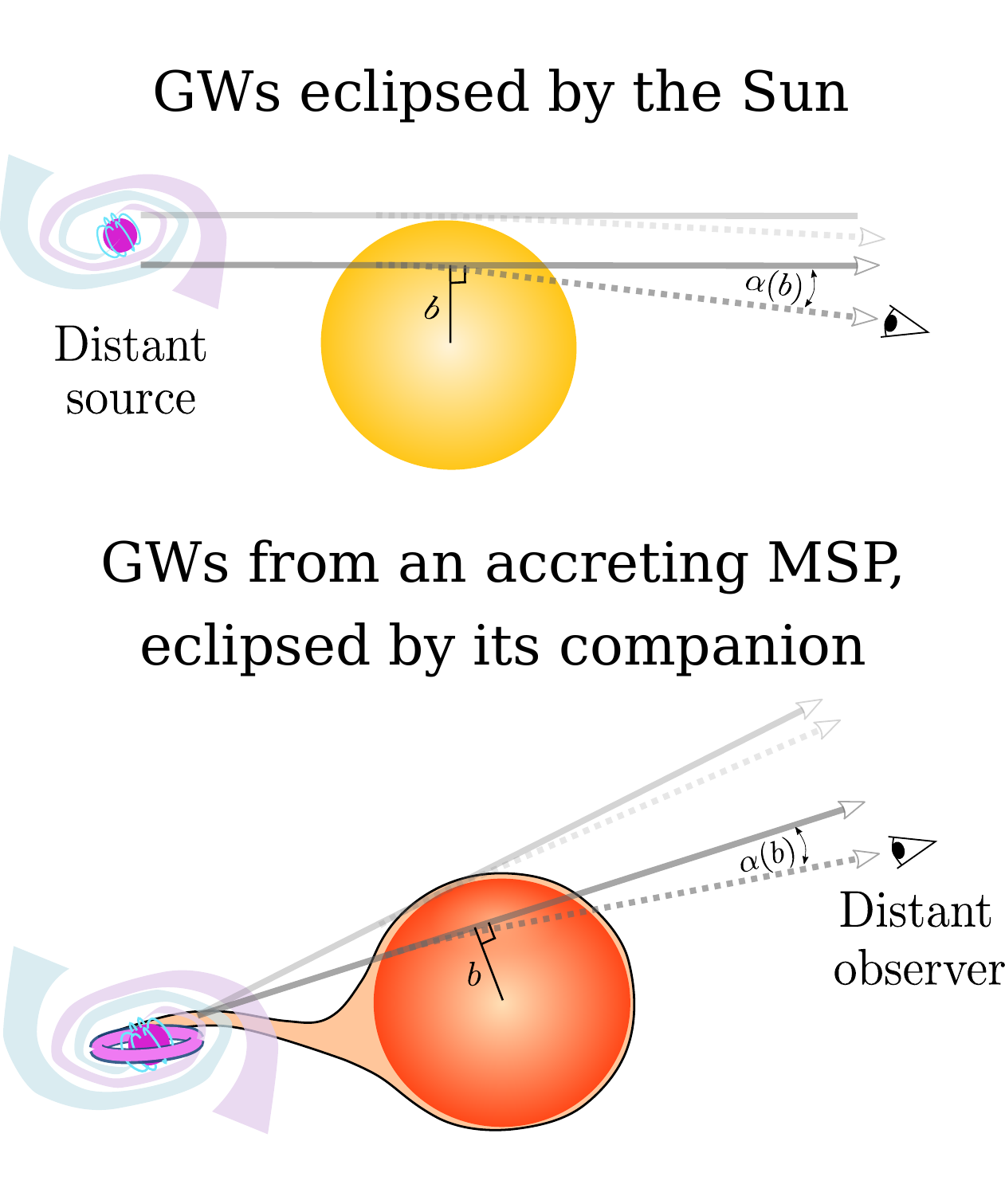}
      \caption{Two different types of eclipses that could allow the measurement
      of the internal mass distribution of stars. (top) Eclipse of
      a distant GW source by the Sun, observed at Earth. In this case, all rays
      approaching the Sun are essentially parallel to each other, resulting in
      caustics far from Earth. (bottom) GWs from an accreting NS eclipsed by its
      companion.}
      \label{fig:sources}.
   \end{center}
\end{figure}

Accreting millisecond X-ray pulsars (AMXPs) with a Roche-lobe filling
companion are expected to be particularly strong sources of high-frequency
CWs. These objects are believed to reach a point where accretion spin-up is
compensated by energy losses from the emission of GWs
\cite{Wagoner1984,Bildsten1998}, which could also result from the  excitation of the
r-mode instability \cite{Haskell2015}.
Moreover, systems
undergoing Roche lobe overflow have a high probability that the compact object
is eclipsed by its companion (see Fig.~\ref{fig:sources} and Fig.~\ref{fig:numsources}).
For a system with a mass ratio $q\equiv M_{\rm
NS}/M_2=10$, the probability that it undergoes eclipses is $\sim 20\%$, and in
particular there is one known eclipsing AMXP, SWIFT J1749.4$-$2807
\cite{MarkwardtStrohmayer2010}.

As depicted in Fig.~\ref{fig:sources}, there are two different situations of interest.
If GWs cross the
Sun, then {by} measuring them we could extract information about the solar interior.  Meanwhile, 
if the eclipsing source is the companion star of an accreting MSP, GWs
provide a probe into the mass distribution of the companion. Eclipses by
the Sun would happen annually and last at most 12 hours, while eclipses from the
companion of an accreting source would happen every orbit, and could last for more
than 10\% of the orbital period depending in the orbital inclination and the
mass ratio.

\begin{table*}
   \centering
   \caption{\label{table:sources} Known pulsars with rotational frequencies
   $>10$ Hz and accreting neutron stars near the ecliptic.}
   \begin{tabular}{lcccc}
      \hline
      \hline
      Source & note & $\nu$ $[\rm Hz]$ & ecliptic latitude & reference \\
      \hline
      J1022+1001 & MSP & 60.8 & $-0.064^\circ$ & \cite{Reardon+2016}\\
      J1730$-$2304 & MSP & 123 & $0.19^\circ$ & \cite{Reardon+2016}\\
      J1142+0119 & MSP & 197 & $-0.58^\circ$ & \cite{Ray+2012}\\
      J1646$-$2142 & MSP & 171 & $0.65^\circ$ & \cite{Ray+2012}\\
      The Crab pulsar  & young pulsar & 29.9& $1.3^\circ$ & \cite{Lyne+2015}\\
      Sco X-1 & accreting NS & - & $5.5^\circ$ & \cite{Bradshaw+1999}\\
      XTE J1751$-$305 & accreting MSP & 435 & $-7.2^\circ$ & \cite{Markwardt+2002}\\
      SWIFT J1749.4$-$2807 & accreting, eclipsing MSP & 518 & $-4.7^\circ$ &
      \cite{Ferrigno+2011}\\
      \hline
   \end{tabular}
\end{table*}

In Table~\ref{table:sources} we summarize a few known sources of interest.
From the ATNF pulsar catalogue \cite{Manchester+2005} we find two recycled
millisecond pulsars (MSPs) that are eclipsed by the Sun, J1022+1001 and J1730$-$2304. MSPs are
stable clocks (cf.\ \cite{Manchester2004}), such that timing of GWs and measurement of the
Shapiro delay \cite{Shapiro1964} during an eclipse might be possible. 
MSPs have low spin-down limits, and Advanced LIGO and Virgo at design sensitivity are not
guaranteed to detect either of these sources \cite{Abbott+2019known_pulsars}.
In contrast, the spin-down
limit has been reached for the Crab pulsar \cite{Aasi+2014}, but it is not eclipsed by the Sun.
Young pulsars also exhibit sudden frequency shifts called glitches
\cite{Espinoza+2011} which make timing of the signal difficult for extended periods
of time. But even for prolific glitchers like the Crab there have not been two
glitches detected less than 10 days apart from each other \cite{Lyne+2015},
making the likelihood of a glitch happening during an eclipse small. Six more
pulsars from the ATNF catalogue are eclipsed by the Sun, but
their low frequencies ($\nu<10~\mathrm{Hz}$) make them unsuitable. 

Sco X-1 and XTE J1751$-$305 are two AMXPs which are close to the
ecliptic, though not close enough to be eclipsed by the Sun.
Searches of the first and second observing runs of Advanced LIGO have provided
upper limits on potential GW emission from Sco X-1
\cite{2017PhRvD..95l2003A,2017ApJ...847...47A,2019arXiv190612040T}, 
and searches of initial LIGO data have provided upper limits for XTE J1751$-$305 \cite{Meadors+2017}.
SWIFT J1749.4$-$2807 is also an AMXP, and a particularly
interesting source because it undergoes periodic eclipses
from its companion. The number of AMXPs has grown
significantly in the last decade \cite{PatrunoWatts2012}.  Although none of
the 19 AMXPs known so far are eclipsed by the Sun, it is likely that an
eclipsing source will be found with further detections. This is particularly
relevant, as AMXPs are a favoured candidate for the first detection of high-frequency GWs \cite{Lasky2015}.

\section{Effects of the Sun on eclipsed GWs}
\label{effects}
We consider the impact of three different effects. As a GW signal passes near the
Sun, it experiences gravitational deflection, which also impacts the
apparent luminosity of the source. In addition, the time of arrival of signals is
delayed compared to what would happen if the Sun was absent, which constitutes
the Shapiro delay \cite{Shapiro1964}.
We first use geometrical optics to compute these effects as observed from the
Earth, and then perform wave optics calculations to check at which frequencies
geometric optics is a good approximation.
We use {a $1M_\odot$ model computed until an age of $4.57\,{\rm
Gyr}$} \cite{Bahcall+1995} with the \texttt{MESA}
\cite{Paxton+2011,Paxton+2013,Paxton+2015,Paxton+2018} code {(version
\texttt{r10398}) to represent the Sun}, and all calculations assume
a radial mass distribution. {Although our model is not
calibrated to constraints from asteroseismic or neutrino measurements of the
Sun, it matches the mass profile $m(r)$ of the calibrated solar models computed by
\cite{Vinyoles+2017} to within $2\%$.}

The Earth is located
far from the caustics produced by the solar lens, and in its neighborhood
the predicted perturbations to the waveform vary on lengthscales of the
order of $R_\odot$. We then expect the geometric optics approximation to apply
for $\lambda \ll R_\odot$, which is the case for rapidly rotating NSs that
emit GWs at frequencies above $1000~\mathrm{Hz}$. 

\subsection{Deflection and amplification}
\begin{figure}
   \begin{center}
      \includegraphics[width=\columnwidth]{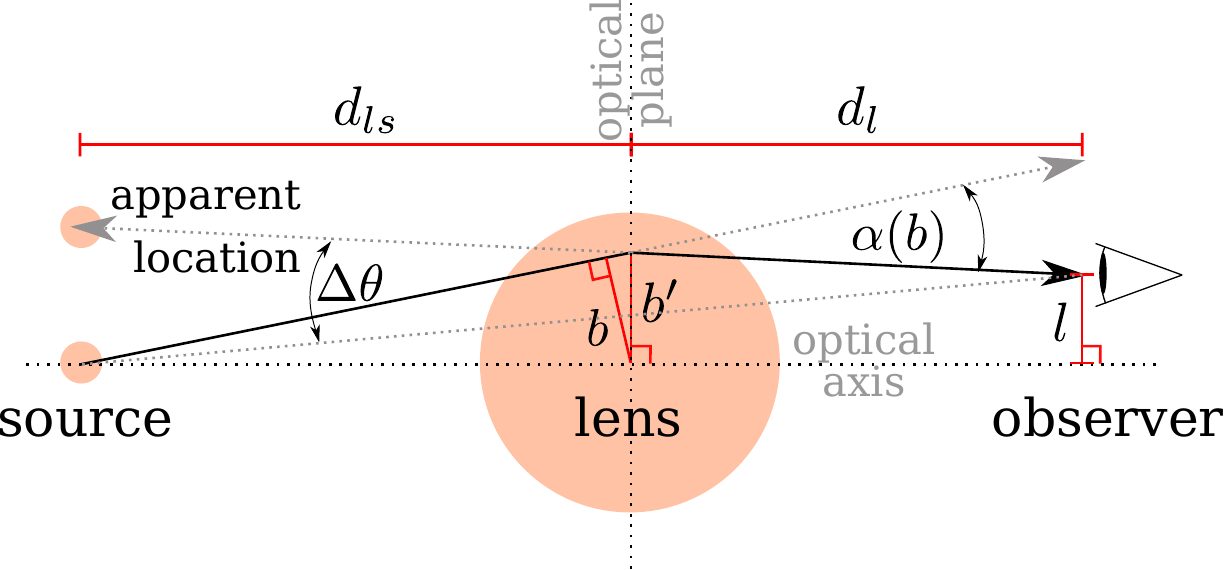}
      \caption{Definition of variables used to compute the change in sky location and amplification of signals from the solar lens. The deflection angle is greatly
      exaggerated.}
      \label{fig:diagramlens}
   \end{center}
\end{figure}

The deflection and amplification produced by a spherically symmetric
gravitational lens are well known results of lensing theory (cf.~\cite{Clark1972}). The deflection
angle can be obtained in terms of the distance of closest
approach $b$, and the mass contained within an infinite cylinder of radius $b$
centered at the lens, $M_{\rm cyl}(b)$. For the Sun, the deflection angle is 
\begin{eqnarray}
   \alpha(b) = \frac{4GM_{\rm cyl}(b)}{c^2 b} = 1.75'' \frac{M_{\rm
   cyl}(b)}{M_\odot}\left(\frac{b}{R_\odot}\right)^{-1},\label{equ::angle}
\end{eqnarray}
{
and $M_{\rm cyl}(b)$ can be computed from a spherically symmetric density
profile $\rho(r)$ and spherical mass coordinate $m(r)$ as
\begin{eqnarray}
   \begin{aligned}
   M_{\rm cyl}(b) = m(b)\qquad\qquad\qquad\qquad\qquad\qquad\qquad \\
   +4\pi\int_b^{R_\odot}\rho(r)r^2\left(1-\sqrt{1-\frac{b^2}{r^2}}\right)
   \mathrm{d}r.
   \end{aligned}
\end{eqnarray}
}
\begin{figure}
   \begin{center}
      \includegraphics[width=\columnwidth]{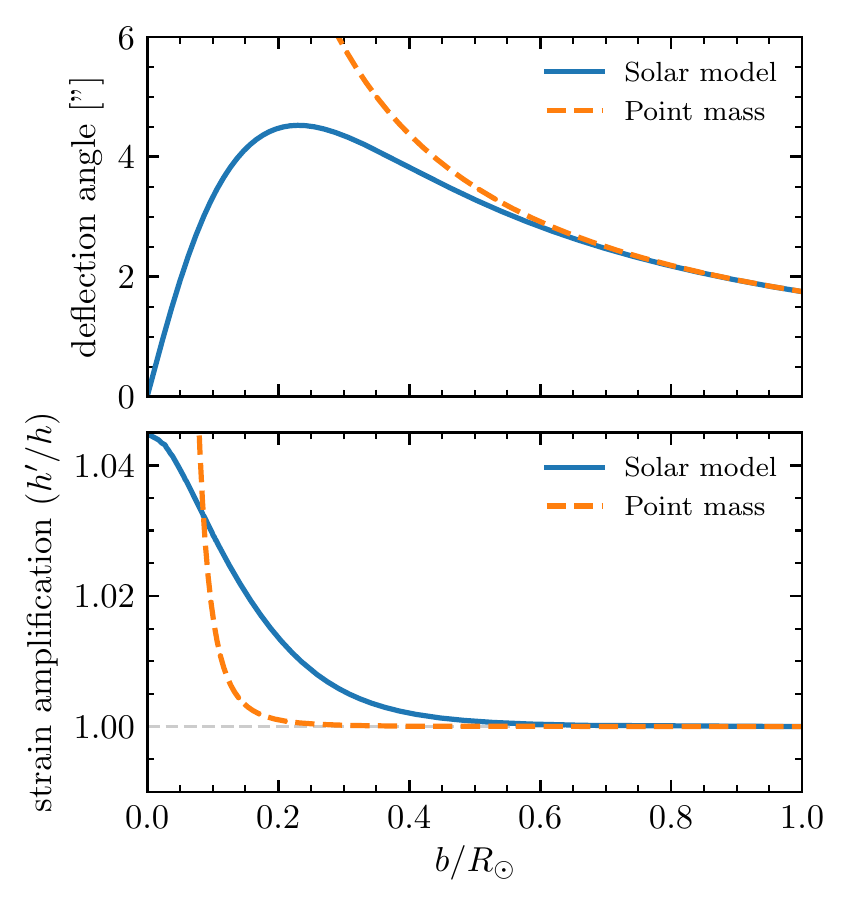}
      \caption{(top) Deflection $\alpha$ of a GW as a function of its
      distance of closest approach to the center of the Sun.
      (bottom)
      Amplification of the GW strain as a function of its
      distance of closest approach to the center of the Sun, considering a
      detector located at the Earth. In both panels, the dashed line indicates the corresponding effect if the Sun were a point mass.}
      \label{fig:angle}
   \end{center}
\end{figure}
Figure~\ref{fig:angle} shows the expected effect from a detailed solar model,
showing that the maximum angle of deflection is smaller than $5''$. An observer
does not measure this angle directly, but instead detects an angular variation
$\Delta \theta$ for the location of the source in the sky. Defining the optical
axis as the line joining the center of the lens and the source, and the optical
plane as the plane perpendicular to the optical axis that crosses the lens, we
approximate the effect of the lens as simply kinking an incoming ray by an angle
$\alpha(b)$ at the optical plane (see Fig.~\ref{fig:diagramlens}). In addition,
$b$ is approximated as the distance of closest approach of the undeflected ray.
This is a standard approximation that is justified when the deflection angle
$\alpha$ is small \cite{Refsdal1964}. Under these assumptions we have that
\begin{eqnarray}
   \Delta \theta = \alpha\sqrt{\frac{d_{ls}^2+b'^2}{l^2+(d_{ls}+d_l)^2}},\label{equ:deflection_rel}
\end{eqnarray}
where we have assumed $\Delta \theta$ and $\alpha$ are small angles, $d_{ls}$ is
the distance between the source and the lens and $d_l$ is the distance between
the source and the observer along the optical axis. The distance between the
center of the lens and the point where the undeflected ray would intersect the
optical plane is denoted by $b'$, while the distance between the observer and
the optical axis is denoted as $l$. The values of $b'$ and $l$ can be computed
as
\begin{eqnarray}
   b'=\frac{b d_{ls}}{\sqrt{d_{ls}^2-b^2}},\qquad l=\frac{d_{ls}+d_l}{d_{ls}}b'-d_l \alpha(b).
\end{eqnarray}
Considering a distant source lensed by the Sun and observed from Earth,
$d_{ls}\gg d_l$ and $d_{ls}\gg l$, such that Eq.~\eqref{equ:deflection_rel}
results in $\Delta \theta\simeq \alpha$ (i.e.\ the deflection angle is equal to
the apparent change in location of the source). Measuring this deflection angle
would provide a direct measurement of the solar
mass distribution \cite{CyranskiLubkin1974}.

The change in strain is equal to the square root of the
change in luminosity.  Following Fig.~\ref{fig:diagramlens} and applying the geometrical optics approximation results
in \cite{Clark1972}
\begin{eqnarray}
   \frac{h'(b)}{h}=\left(\frac{d_{ls}+d_l}{d_{ls}}\right)\sqrt{\frac{b'}{l}}\left(\frac{d_{ls}+d_l}{d_{ls}}-d_l\frac{\mathrm{d}\alpha}{\mathrm{d}b}\right)^{-1/2},
   \label{equ::amp}
\end{eqnarray}
where $h$ is the strain that would be measured in the absence of lensing.
For the case of a distant source being lensed by the Sun $d_{ls}\gg d_l$ and $b'$
is almost equal to $b$, leading to
\begin{eqnarray}
   \frac{h'(b)}{h}=\left(\frac{b}{b-a_\oplus
   \alpha(b)}\right)^{1/2}\left(1-a_{\oplus}\frac{\mathrm{d}\alpha}{\mathrm{d} b}\right)^{-1/2},\label{equ::ampsun}
\end{eqnarray}
where $d_l=a_\oplus=1\;\rm au$. The result from our solar model is shown in
Fig.~\ref{fig:angle}.
The maximum amplification, which happens at the core of the Sun, is
a factor {$1.043$} of the non-lensed strain. In contrast, a point mass results in
no amplification during most of an eclipse, with a steep rise below
$b/R_\odot\sim 0.1$. The large deflection angles that would be
produced by a point mass result in rays with $b/R_\odot\sim 0.05$ focusing at
$1$ au from the Sun, at which point the geometric optics approximation is not
valid. 

\subsection{Shapiro delay}
\begin{figure}
   \begin{center}
      \includegraphics[width=\columnwidth]{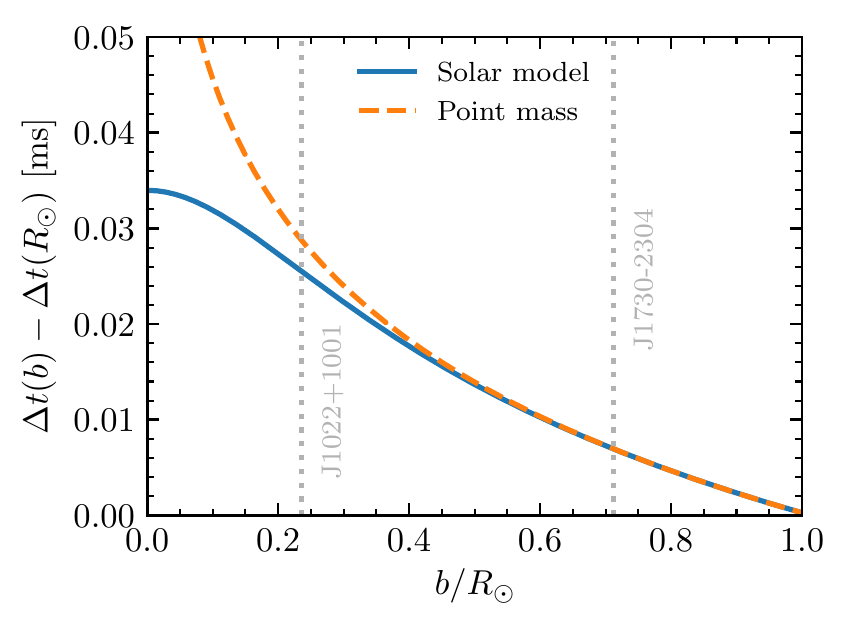}
      \caption{The time delay imprinted on a GW as a function of its
      distance of closest approach to the center of the Sun. The solid line
      shows the delay derived from a detailed solar model computed with the
      \texttt{MESA} code, while the dashed line is the effect from a point mass
      and represents an absolute upper limit. The two dotted
      vertical lines indicate the lowest value of $b$ reached by the two known
      MSPs that are eclipsed by the Sun.}
      \label{fig:timedelay}
   \end{center}
\end{figure}

The time delay of a signal as it passes by the Sun,
compared to the arrival time the signal would have in the absence of the Sun, is given by
\cite{Shapiro1964,BackerHellings1986}
\begin{eqnarray}
   \Delta t=-\frac{2}{c^3}\int_A^B \Phi \,\mathrm{d}l, \label{eq:intline}
\end{eqnarray}
where $A$ and $B$ denote the location of the source and the receiver, and $\Phi$ is the gravitational potential of the Sun which satisfies Poisson's equation $\nabla^2 \Phi=4\pi G\rho$ and, assuming spherical symmetry for the Sun, is given by $\Phi=-GM_\odot/r$ for $r>R_\odot$.  Equation~\eqref{eq:intline} is given in coordinate time, and the actual time delay measured on Earth includes additional small corrections that depend on the Solar System ephemerides
\cite{BackerHellings1986}. 
The integral in Eq.~\eqref{eq:intline}
can be estimated by integrating through the straight line path that the
unperturbed light ray would follow, as the additional delay produced by the
deflection of the null geodesic only adds up to a few tens of nanoseconds
\cite{RichterMatzner1983}. If the trajectory does not go through the Sun, then
Eq.~\eqref{eq:intline} is equal to
\begin{eqnarray}
   \Delta t_{\rm out} =-\frac{2GM_\odot}{c^3}\ln\left(\frac{\hat{n}\cdot
   \vec{a}_{\oplus}+a_{\oplus}}{\hat{n}\cdot \vec{a}_{\rm s}+a_{\rm s}}\right),
\end{eqnarray}
where $\vec{a}_{\rm s}$ and $\vec{a}_{\rm \oplus}$ are the positions of the source
and the receiver with respect to the Sun, and $\hat{n}$ is a unit vector from
the receiver to the source. If the source is far away, such that
$\hat{n}\cdot \vec{a}_{\rm s}+a_{\rm s}\simeq 2a_{\rm s}$, the time delay can be
approximated as
\begin{eqnarray}
   \Delta t_{\rm out} = -\frac{2GM_\odot}{c^3}\ln\left[\frac{a_{\oplus}}{2a_{\rm
   s}}(1-\cos\beta)\right],\label{eq:beta}
\end{eqnarray}
where $\beta$ is the angle in the sky between the center of the Sun
and the source, {as observed from the location of the receiver}
($\cos\beta=-\hat{n}\cdot\vec{a}_\oplus/a_\oplus $). For sources
close to the solar disk $\beta$ is small, such that
the time delay can be expressed in terms of the distance of closest approach to
the Sun $b\simeq \beta a_{\oplus}$,
\begin{eqnarray}
   \Delta t_{\rm out}{(b)} =
   -\frac{4GM_\odot}{c^3}\ln\left(\frac{b}{2\sqrt{a_{\oplus}a_{\rm
   s}}}\right).\label{eq:dt0}
\end{eqnarray}
If the line does go through the Sun, then this equation has to be corrected
for the part of the trajectory that crosses it,
\begin{eqnarray}
   \Delta t{(b)} &=&\Delta t_{\rm out}{(b)} - \Delta t_{-}{(b)} + \Delta t_{+}{(b)}, \\
   \Delta t_{-}{(b)} &=& \frac{2}{c^3}\int_{A'}^{B'} \frac{GM_\odot}{r} \,\mathrm{d}l, \\
   \Delta t_{+}{(b)} &=& {-}\frac{2}{c^3}\int_{A'}^{B'} \Phi(r) \,\mathrm{d}l,\label{eq:dt1}
\end{eqnarray}
where $A'$ and $B'$ are the points where the trajectory crosses the surface of
the Sun. Computing $\Delta t_{-}$ yields
\begin{eqnarray}
   \Delta t_{-}{(b)} =
   -\frac{4GM_\odot}{c^3}\ln\left(\frac{b}{R_\odot+\sqrt{R_\odot^2-b^2}}\right),\label{eq:dt2}
\end{eqnarray}
while $\Delta t_{+}$ can be transformed into an integral over the mass
coordinate of the Sun,
\begin{eqnarray}
   \Delta t_+{(b)} &=&
   -\frac{4}{c^3}\int_b^{R_\odot}\frac{\Phi(r)r}{\sqrt{r^2-b^2}}\,\mathrm{d}r \nonumber \\
   &=&
   \frac{4GM_\odot}{c^3}\frac{\sqrt{R_\odot^2-b^2}}{R_\odot} \nonumber \\
   && + \left. \frac{4}{c^3}\int_b^{R_\odot}\frac{Gm(r)}{r^2}\sqrt{r^2-b^2}\,\mathrm{d}r \right. \label{eq:dt3},
\end{eqnarray}
where we have used $\mathrm{d} \Phi/\mathrm{d} r = Gm(r)/r^2$ and
$\Phi(R_\odot)=-GM_\odot/R_\odot$.
Since only relative changes in the arrival time of pulses can be determined, it
is more useful to consider the difference between the delay time for $b$,
and the delay time of a signal that passes right by the surface of the
Sun ($b=R_\odot$). Combining Eq.~\eqref{eq:dt0}, \eqref{eq:dt1}, \eqref{eq:dt2} and \eqref{eq:dt3} then gives us
the time delay as a function of $b$ and the mass profile of the Sun $m(r)$,
\begin{eqnarray}
   \Delta t(b)-\Delta t(R_\odot)=-\frac{4GM_\odot}{c^3}\qquad\qquad\qquad\qquad \nonumber \\
      \times\left[\ln\left(
      \frac{R_\odot+\sqrt{R_\odot^2-b^2}}{R_\odot}\right)-\frac{\sqrt{R_\odot^2-b^2}}{R_\odot}\right. \nonumber  \\
   \left.-\frac{1}{M_\odot}\int_b^{R_\odot}\frac{m(r)}{r^2}\sqrt{r^2-b^2}\,\mathrm{d}r\right].
\end{eqnarray}
This can be rewritten in a way that clearly distinguishes the contribution for
the case of a point mass,
\begin{eqnarray}
   \Delta t(b)-\Delta t(R_\odot)=-\frac{4GM_\odot}{c^3}\left[\ln\left(
   \frac{b}{R_\odot}\right)\right.\qquad\qquad \nonumber \\
   \left.{+}\frac{1}{M_\odot}\int_b^{R_\odot}\frac{M_\odot-m(r)}{r^2}\sqrt{r^2-b^2}\,\mathrm{d}r\right].
   \label{eq:delaysol}
\end{eqnarray}
The factor $4GM_\odot/c^3=0.02~\mathrm{ms}$ shows how small the expected effect is.
The time delay is plotted in Fig.~\ref{fig:timedelay}, and the largest delay is
of {$\sim
0.034~\mathrm{ms}$} for a source crossing the center of the Sun. This represents
a shift in the phase of the pulsars listed in Table~\ref{table:sources} ranging from a half of a percent to a few
percent. When the source is not eclipsed the Shapiro delay still changes
depending on the angle $\beta$ between the locations of the source and the Sun
in the sky. Combining Eq.~\eqref{eq:beta} and Eq.~\eqref{eq:dt0} and considering
a source {located on the opposite side of the sky from the Sun} ($\beta=\pi$) results in
\begin{eqnarray}
\begin{aligned}
   \Delta t(\beta=\pi) - \Delta t(R_\odot)&=-\frac{4GM_\odot}{c^3}\ln\left(\frac{2a_\oplus}{R_\odot}\right)\\
   &=-0.12~\mathrm{ms}.\label{equ:noeclipse}
   \end{aligned}
\end{eqnarray}
The magnitude of this orbital variation in the time delay is larger than that during an eclipse, but it can only provide information on the total mass of the Sun rather than its internal structure.

Equation~\eqref{eq:delaysol} provides information on the solar interior in the form of
an integral over the mass distribution. If the derivative of the time delay as a
function of $b$ can be measured as the source passes behind the Sun, it provides a direct measurement of $M_{\rm
cyl}(b)$,
\begin{eqnarray}
   \frac{\mathrm{d}\Delta t}{\mathrm{d}b} = -\frac{4GM_{\rm cyl}(b)}{c^3b}= -\frac{\alpha(b)}{c}.
\end{eqnarray}
This relation between the deflection angle and the time delay is exactly what is
expected in terms of the change in direction of propagation of an incoming
wavefront.


\subsection{Wave optics}
\label{sec:wave}

Calculations using geometric optics are only valid in the limit that the
effects of the lens on the amplitude, phase, and direction of propagation of a wave
occur on lengthscales much larger than a wavelength, and on timescales much
longer than the period of the wave. For the case of an eclipse by the Sun being
observed at Earth, all predicted effects on incoming waves are small and operate on a
lengthscale $\sim R_\odot$, such that the geometric optics approximation
requires $\lambda \ll R_\odot$. The timescale on which the properties of the
wave change corresponds to the {duration} of the eclipse $\tau_{\rm ec}$, which sets a limit on the
frequency of the source for the applicability of geometric optics, $f_{\rm
GW}^{-1}\ll\tau_{\rm ec}$. The tighter constraint is provided by the limit on the
wavelength; for $\lambda=R_\odot$ the corresponding frequency is $f_{\rm GW}\sim
0.5~\mathrm{Hz}$, so the geometric optics approximation requires $f_{\rm GW}\gg
0.5~\mathrm{Hz}$. {This implies that our geometric optics calculations
are only applicable in the high-frequency range that is probed by ground based
observatories, while at lower frequencies the impact of wave optics needs to be
analyzed with care.}

For practical purposes, it is necessary to quantify how much
smaller than $R_\odot$ the wavelength needs to be for {wave optics effects to become negligible}. For this, we need to drop the assumption of geometric
optics. Following \cite{Ohanian1974,BontzHaugan1981}, we compute the Kirchhoff
integral, which allows the calculation of a wave given
its properties on a surface surrounding the observation point. The effect of the
lens is encoded by the time delay given by Eq.~\eqref{eq:delaysol}, which
produces a phase shift at the lens plane. Given this, the Kirchhoff integral can
be numerically computed to determine the amplification and the time delay
observable at any point in space (see Appendix~\ref{ap:k}).

\begin{figure}
   \begin{center}
      \includegraphics[width=\columnwidth]{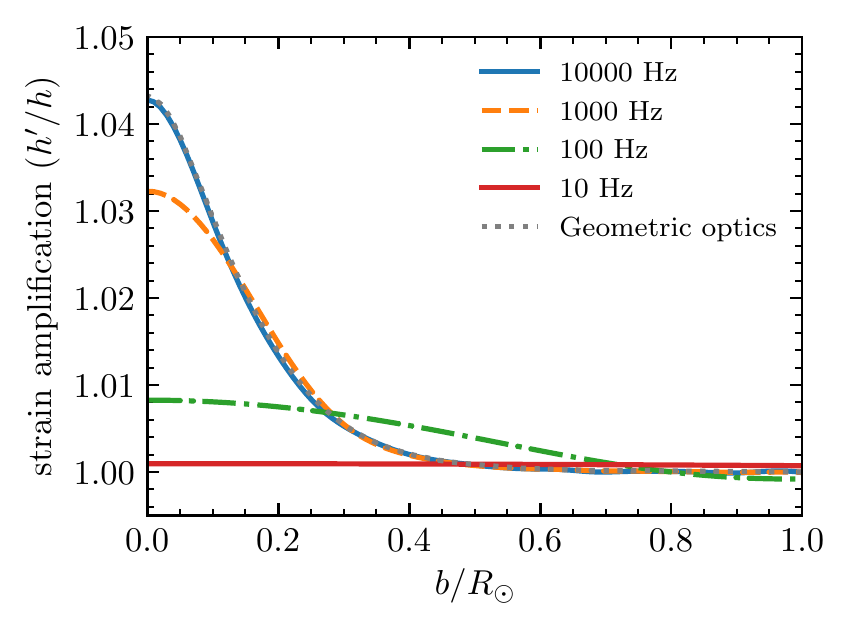}
      \caption{
      Amplification of the GW strain observable at Earth during an eclipse.
   Each line indicates a different frequency, while the geometric optics result
      is the expected value in the limit that $f_{\rm GW}\rightarrow\infty$.}
      \label{fig:woamp}
   \end{center}
\end{figure}

\begin{figure}
   \begin{center}
      \includegraphics[width=\columnwidth]{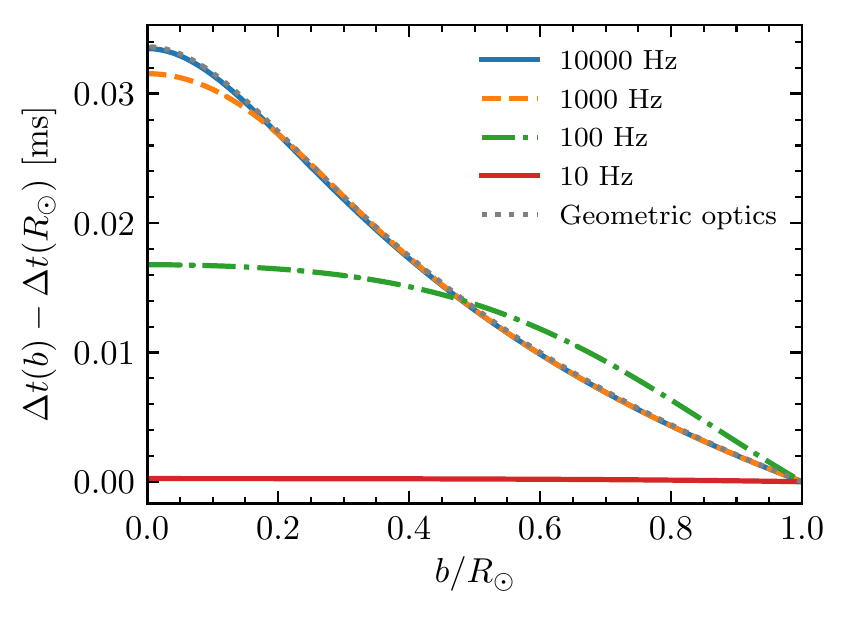}
      \caption{
      Time delay observable at Earth during an eclipse.
   Each line indicates a different GW frequency; the geometric optics result
      is the expected value in the limit that $f_{\rm GW}\rightarrow\infty$.}
      \label{fig:wodelay}
   \end{center}
\end{figure}

Using our solar model, we compute the amplitude and time delays
observable at Earth through an eclipse for sources at different frequencies,
which we show together with the expected results from geometric optics in
Fig.~\ref{fig:woamp} and \ref{fig:wodelay}. As expected, at frequencies below
10 Hz the amplification and the time delay are negligible, meaning that
waves with that frequency are unaffected by the solar lens. At $100~\mathrm{Hz}$ the
{solar lens can amplify a signal by up to $\sim 1\%$, and delay it by
$\sim0.02\,{\rm ms}$, but it still deviates significantly from the geometric optics
calculation.}
At $10^3~\mathrm{Hz}$, the effects predicted using wave optics
closely match those of geometric optics except for the inner $10\%$ of the Sun,
with the amplification and time delays for a source passing through the very center of the Sun
being $\sim30\%$ and $5\%$ lower {than} the results of geometric optics. At $10^4~\mathrm{Hz}$,
the predicted effects become almost equivalent to those of geometric
optics; {however, NSs are not expected to emit CWs at or above
$10^4~\mathrm{Hz}$, as their break-up frequencies are expected to be $<3\times
10^3~\mathrm{Hz}$ \cite{Cook+1994} and the fastest spinning known MSP has a
rotation frequency of $716~\mathrm{Hz}$ \cite{Hessels+2006}.}

These results show that the ideal signals to extract information about the solar
interior are GWs with $f_{\rm GW}\ge 10^3~\mathrm{Hz}$, as at these
frequencies the amplitude of the predicted effects is almost maximal, and having
results close to the geometric optics prediction makes the inverse problem of
deducing the structure of the Sun from the signal easier. This has to
be put in contrast with the results of \cite{Ohanian1973, Ohanian1974,
BontzHaugan1981}, who determined that no significant amplification can happen
for waves with frequencies $f_{\rm GW}<10^4~\mathrm{Hz}$. The main difference with our
work is that they were considering amplification at the caustics of the Sun,
regions in space where multiple images are formed and geometric optics
predicts infinite amplification. The situation we are studying is significantly
different, as the Earth is located far away from a caustic. From our computed
solar model, the nearest caustics to the Sun are at a distance of $\sim20~\mathrm{au}$
from it, near the orbit of Uranus. In contrast to that previous work, we find that the geometric optics limit is fully recovered for $10^4~\mathrm{Hz}$ signals observed at the
Earth.

Despite our expectation that CBCs occurring behind the Sun are extremely
uncommon, if one happens right behind the center
of the Sun it would experience an anomalous increase in amplitude of a few
percent. This is because as the signal chirps to higher frequencies, the
predicted amplitude will approach the expected result from geometric optics.

\section{GWs in accreting Neutron Stars eclipsed by a binary companion}
\label{sec:eclipse_binary}
When the eclipse is produced by a nearby binary companion of the GW source we have that the orbital separation $a=d_{ls} \ll d_l$ (see Fig.~\ref{fig:diagramlens}) and Eq.~\eqref{equ::amp} yields
\begin{eqnarray}
   \frac{h'(b)}{h}=\left(\frac{b'}{b'-a
   \alpha(b)}\right)^{1/2}\left(1-a\frac{\mathrm{d}\alpha}{\mathrm{d} b}\right)^{-1/2}.
\end{eqnarray}
This is equivalent to Eq.~\eqref{equ::ampsun} for sources eclipsed by the Sun, except that the distance between the Earth and the Sun $a_\oplus$ is replaced by the orbital separation $a$, and $b$ is switched for $b'$ as the approximation $b\simeq b'$ is no longer valid. If we consider the eclipsing object to be a Roche-lobe filling star similar to the Sun, then $a\sim R_\odot\ll a_\oplus$, resulting in a much smaller amplification than when distant sources observed from the Earth are eclipsed by the Sun. The angle of deflection $\alpha$ in this case is computed in the same way as for a source eclipsed by the Sun, but the apparent change in location $\Delta \theta$ is much smaller; in the limit $d_l\gg d_{ls}$ Eq.~\eqref{equ:deflection_rel} results in
\begin{eqnarray}
   \Delta \theta = \frac{d_{ls}}{d_l}\alpha.
\end{eqnarray}
Thus, we do not expect amplification or deflection to be relevant when these sources are observed using GW detectors.

However, the magnitude of the Shapiro delay that would be measurable at the Earth is the same
if one considers an eclipsing Sun-like companion star as for the case of
eclipsing by the Sun. The derivation is completely analogous to the one in the
previous section, except that the position of the detector and the GW source are
inverted. For the case of an edge-on system, Eq.~\eqref{eq:beta} is also valid,
with $\beta$ corresponding to the angle in the sky between Earth and the binary
companion, as observed from the GW source. The orbital phase is then equal to
$\beta/2\pi$, and during an eclipse the small angle approximation $b=\beta a$ is
still valid, where $a$ is the orbital separation.
{This means Eq.~\eqref{eq:delaysol} can be used to
compute the expected time delay during an eclipse for a given impact parameter
$b$.}

To evaluate the magnitude of this time delay during a mass transfer phase we use
the \texttt{MESA} code to model a low-mass X-ray binary consisting of a
$1.4M_\odot$ NS and 
a $1M_\odot$ zero-age main sequence stellar companion with an initial orbital period of $2$ days.  We
account for magnetic braking as in \cite{Rappaport+1983}, which efficiently
removes orbital angular momentum from the system. This leads to Roche lobe
overflow when the system is $\sim6~\mathrm{Gyr}$ old and the orbital period is 
$0.4$ days, at which point the star is similar to our Sun. Further loss of
orbital angular momentum due to magnetic braking keeps shrinking the orbit and
reduces the orbital period to $1.4\; \rm hr$, while the mass of the donor star
decreases to $0.1 M_\odot$
through mass transfer in $\sim 2.5~\mathrm{Gyr}$.

\begin{figure}
   \begin{center}
      \includegraphics[width=\columnwidth]{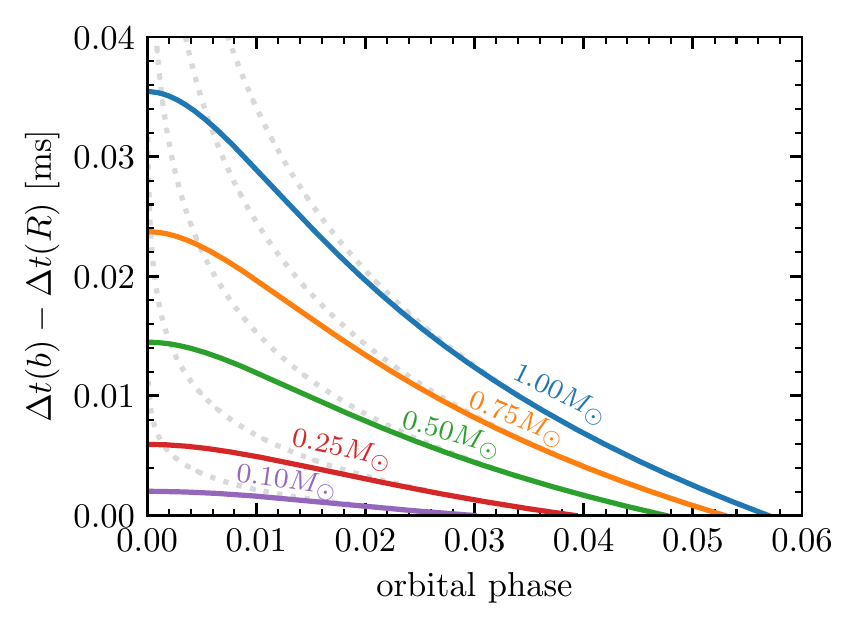}
      \caption{Time delay as a function of orbital phase for a system consisting of a
      Roche-lobe-overflowing donor star eclipsing a $1.4M_\odot$ NS companion.
      The donor is initially a $1\, M_\odot$ star at solar metallicity in a
      binary with an initial orbital period of $2$ days. Each solid line
      corresponds to a different point in time as mass transfer reduces the mass
      of the donor star to the value appearing in the caption, and shows the
      expected time delay from geometric optics for an edge-on system. Dotted
      lines indicate the expected time delay for the case of a point mass rather
      than a donor star model with an extended mass distribution.}
      \label{fig:timedelay2}
   \end{center}
\end{figure}

Figure~\ref{fig:timedelay2} shows the expected Shapiro time delay as a function of the
orbital phase, in case this system is observed edge-on. Eclipses last
for more than $10\%$ of the orbital period at the beginning of mass transfer,
and the expected time delay is the same as the one we computed for the Sun in
the previous section. As mass transfer proceeds, the magnitude of the time delay
decreases from a few tens of microseconds down to just $2$ microseconds, and the
duration of the eclipses decrease as well.  During mass transfer the
amplification of the strain is always below $0.07\%$, which is almost two orders
of magnitude smaller than for sources eclipsed by the Sun and observed from
Earth.

A GW measurement of the Shapiro time delay from the eclipsing companion will
also provide an independent estimate {of} the properties of the system even if the
resolution is insufficient to probe the companion's internal mass distribution.
Coupled with the known inclination -- which is constrained to be near edge-on by
virtue of observing the eclipse altogether -- and a measurement of the radial
velocity variation of the GW source, the Shapiro time delay breaks the usual
mass function degeneracy in X-ray binaries, allowing the NS mass to be inferred.
For this purpose the change in the Shapiro delay through an entire orbital phase
can be used. Similarly to Eq.~\eqref{equ:noeclipse}, for an edge-on system
observed at the point where the GW source is in front of the star, the relative
delay compared to the point where $b=R$ is
\begin{eqnarray}
   \Delta t(\beta=\pi) - \Delta t(R_\odot)=-\frac{4GM}{c^3}\ln\left(\frac{2a}{R}\right)\label{equ:noeclipse_bin},
\end{eqnarray}
where $M$ and $R$ are the mass and radius of the eclipsing star, respectively.
For a Roche-lobe filling star identical to the Sun with a $1.4M_\odot$
companion, the orbital separation is $a=2.9 R$, and
Eq.~\eqref{equ:noeclipse_bin} gives a relative time delay of $-0.034$~ms,
essentially doubling the effect observable just during an eclipse. However, for
AMXPs X-ray timing may provide a better tool to measure the Shapiro delay (cf.\
\cite{MarkwardtStrohmayer2010}). Moreover, if radial velocity measurements of
the companion star are available along with NS radial velocity measurements and
a known inclination from eclipses, the masses can be inferred directly.

\subsection{Number of eclipsed accreting neutron star binaries}
We estimate the number of rapidly rotating NSs that are eclipsed either by a
binary companion or the Sun through a population synthesis of NS binaries using
the binary population synthesis code
\texttt{COSMIC} \cite{Breivik+2019}. 
\texttt{COSMIC} evolves binary
systems with a modified version of the binary evolution code \texttt{BSE}
\cite{Hurley+2002}. The modified version includes updates to account for
metallicity dependent winds \cite{Vink+2001, VinkdeKoter2005}, neutrino driven
core collapse supernova explosions \cite{Fryer2012}, and compact object natal
kicks \cite{Hobbs+2005}. 
We treat the star
formation history (SFH) for the Milky Way Thin Disk, Thick Disk, and Bulge
populations separately as outlined in Table~\ref{table:sfh}. All binaries are
initialized according to the observationally derived correlated joint probability distribution of \cite{Moe2017} over the primary mass, secondary mass, orbital
period, eccentricity, and multiplicity of each binary. We assume all systems with a
multiplicity greater than one are binary systems, thus ignoring triples and higher multiplicity systems. 

\begin{table}
   \centering
   \caption{\label{table:sfh} Star formation history (SFH) for Galactic components following \cite{McMillan2011}.} 
   \begin{tabular}{lcccr}
      \hline
      \hline
      Component & Age [Gyr] & SFH & $Z$ [$Z_{\odot}$] & Mass [$M_{\odot}$] \\
      \hline
      Thin Disk & $10$ & constant & 1 & $4.32\times10^{10}$\\
      Thick Disk & $11$ & $1\,\rm{Gyr}$ burst & 0.15 & $1.44\times10^{10}$ \\
      Bulge & $10$ & $1\,\rm{Gyr}$ burst & $1$ & $8.9\times10^{9}$ \\
      \hline
   \end{tabular}
\end{table}

\begin{figure}
    \centering
    \includegraphics{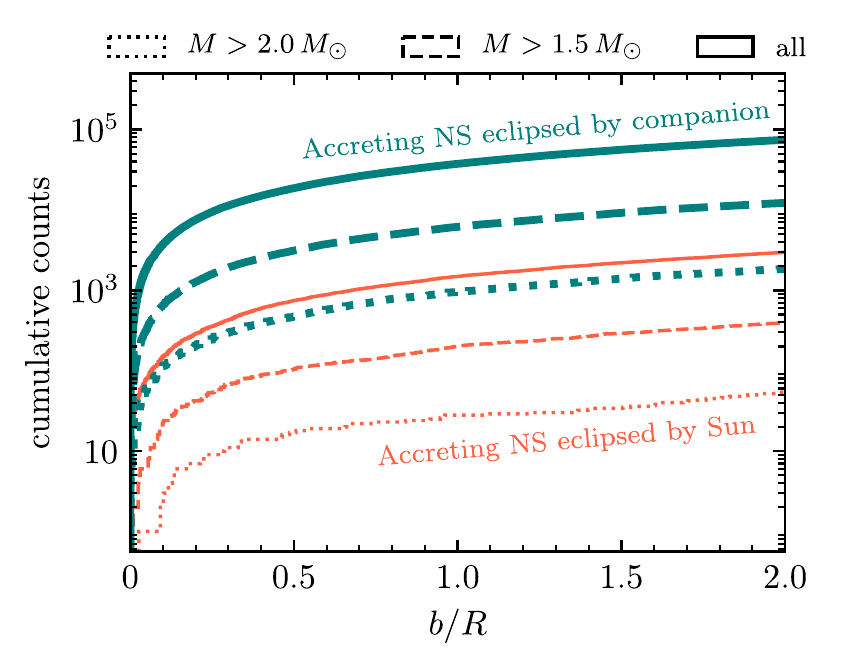}
    \caption{Cumulative counts of impact parameters below the abscissa value for the population of Galactic
    accreting NS binaries eclipsed by a companion (green, in units of donor Roche lobe radius) or the Sun (orange, in solar radii).
    Line styles indicate cuts on the donor mass.}
    \label{fig:NS_pop}
\end{figure}

We simulate all binaries from the zero-age main sequence and restrict our attention
to the population containing a NS that is accreting from a stellar donor with
$M_{\rm{D}}>0.05~\rm{M_{\odot}}$ at present. 
This criterion removes NSs that would experience time delays on the order of a microsecond or less when eclipsed by their companions; 
black widow pulsars, which are observed to eclipse, fall into this category \cite{Fruchter+1988,Lyne+1990,Freire2005,Polzin+2018,Guillemot+2019} and so produce sub-microsecond time delays. 
{We confirm that the present day orbital parameters are well sampled by increasing the number of simulated systems until the binary parameter distributions do not depend on the sample size, as described in \cite{Breivik+2019}}.

We generate $500$ simulated Milky Way populations by re-sampling the simulated
population with replacement.  The number of accreting NS binaries in each Galactic 
component population is found by multiplying the number of simulated binaries by
the ratio of the Galactic component mass to the total mass required to generate
the simulated accreting NS population. Every re-sampled binary is assigned a
position based on its Galactic component distribution following
\cite{McMillan2011} and a random inclination $i$ that is uniform in $\cos{i}$.
 
The statistics from 500 Milky Way populations are summarized in 
Table~\ref{table:popstats}. To check the validity of our Milky Way population model,
we compare to previous population synthesis studies. For our Thin Disk 
population, we first consider \cite{Nelemans2001}, which finds a total population of 
$2.2\times10^6$ NS--white dwarf binaries from a Thin Disk population with total mass 
$\sim6\times10^{10}\,M_{\odot}$ formed at an exponentially decreasing star 
formation rate. We find a total population $5.9\times10^6$ NS-white dwarf binaries, 
without constraints on the donor mass or Roche lobe filling factor.  Our yield is a factor of $\sim4$ greater than \cite{Nelemans2001} if we take into account our lower Thin Disk mass 
of $4.32\times10^{10}\,M_{\odot}$. For the Bulge population, we compare to 
\cite{vanHaaften2015}, which simulated the population of low mass X-ray binaries 
(LMXBs) with NS accretors in the Bulge and predict a population of $\sim2.1\times10^3$ 
NS LMXBs for a total Bulge mass of $1\times10^{10}\,M_{\odot}$. Our model 
roughly agrees with \cite{vanHaaften2015}, though we predict a twice greater yield of NS 
LMXBs once we account for our relatively lower Bulge mass of $8.9\times10^9\,M_{\odot}$. Direct comparisons between population synthesis studies which use different codes is difficult, thus rate differences within an order of magnitude are commonly accepted \cite{Toonen2014}. Scaling our populations numbers down to match the numbers reported by \cite{Nelemans2001} and \cite{vanHaaften2015}, does not change our general conclusions.

\begin{table}
   \centering
   \caption{\label{table:popstats} Accreting NS population statistics for the Galaxy. Here $N_\mathrm{MW}$ denotes the total number of accreting NSs expected in the galaxy with donors $>0.05M_\odot$, $N_{\mathrm{Donor,ecl}}$ is the number of those that are eclipsed by their companions, and $N_{{\odot,\mathrm{ecl}}}$ is the number of those expected to be eclipsed by the Sun.}
   \begin{tabular}{lccr}
      \hline
      \hline
      Component & $N_{\mathrm{MW}}$ & $N_\mathrm{{Donor,ecl}}$ & $N_{{\odot,\mathrm{ecl}}}$\\
      \hline
      Thin Disk & $109,000$ & $36,000\pm160$ & $1500\pm40$ \\
      Thick Disk & $1700$ & $450\pm20$ & $20\pm5$ \\
      Bulge & $3400$ & $920\pm30$ & $30\pm5$ \\
      \hline
   \end{tabular}
\end{table}

For each accreting NS we consider eclipses both by the donor companion and by the Sun. The impact parameter for the eclipse is
\begin{equation}
\label{eq:impact}
    \frac{b}{R} = \frac{a}{R}\,\sin\left(\frac{\pi}{2}-i\right), 
\end{equation}
where for donor eclipses $a$ is the binary semimajor axis, $R$ is the donor
Roche lobe radius, and $i$ is the binary inclination; for solar eclipses $a$ is
an astronomical unit, $R$ is $R_{\odot}$ and $\pi/2-i$ is the ecliptic latitude. 

Figure~\ref{fig:NS_pop} shows the cumulative counts of impact parameters smaller
than a given value for the populations of accreting NSs eclipsed by their donors
(green) and by the Sun (orange) for a single Milky Way population. The different
line styles show the fractions of the population which satisfy donor mass cuts
of $M > 2.0\,M_{\odot}$ and $M > 1.5\,M_{\odot}$. As expected from
Fig.~\ref{fig:numsources}, we find nearly two orders of magnitude fewer solar
eclipses than donor eclipses for our accreting NS population. We find a slight
excess in the fraction of solar eclipses when compared with the probability in
Fig.~\ref{fig:numsources} because NS accretors are highly concentrated in the
Galactic plane, which intersects with the ecliptic plane.

\section{Detectability}\label{sec:detect}

We consider in turn all three effects caused by an eclipse: 
magnification $\Delta h / h\equiv 1-h'/h$, deflection in the apparent sky location due to lensing $\Delta \theta$, 
and the variation in the Shapiro time delay $\Delta t_s\equiv \Delta t(b)-\Delta t(R_\odot)$, as described in Eq.~\eqref{equ::amp}, Eq.~\eqref{equ:deflection_rel}, and Eq.~\eqref{eq:delaysol}, respectively.
As shown in Section~\ref{effects}, the characteristic magnitude of these effects for sources eclipsed by the Sun is
\begin{eqnarray}
\frac{\Delta h}{h} &\sim& 0.01\, ;\\
\Delta \theta &\sim & 1'' \, ; \\
\Delta t_s&\sim & 0.01~\mathrm{ms} \, .
\end{eqnarray}
As discussed in Section~\ref{sec:eclipse_binary}, in the case of a NS emitting
GWs that is eclipsed by a binary companion, both the amplification and
deflection are negligible, but the Shapiro time delay has a comparable magnitude as long as the mass of the companion star is comparable to that of the Sun.

Since the eclipse occupies only a small fraction of the overall observation, 
the measurement uncertainties from comparing the amplitude, location, and timing of the signal
during the eclipse against the non-eclipse values are dominated by the uncertainties during the eclipse.  
Let $\rho$ be the total signal-to-noise ratio (SNR) over the full observation, 
and $\rho_\mathrm{ec}$ the SNR during the eclipse.  
We assume that the SNR is proportional to the square root of the duration of the observation \cite{Jaranowski1998}. 
The scaling is weaker for initial detectability with a semi-coherent search \cite{Wette2012}, but we can assume that the
source has already been detected, since that requires a much lower SNR than is necessary for measuring
the eclipse properties, and a fully coherent analysis is used for parameter estimation.  
For example, for eclipses by the Sun, 
$\rho_\mathrm{ec} \approx \rho (R_\odot/(\pi~\mathrm{au}))^{1/2} 
\approx \rho (12~\mathrm{hr} / \mathrm{yr})^{1/2} \approx 0.04 \rho$.  The SNR
required for measuring the variation of the different effects over the course of
an eclipse can be estimated by dividing the observation into shorter segments.
For example, one-hour observations would have individual SNRs of $\sim
\rho(1~\mathrm{hr} / \mathrm{yr})^{1/2} \approx 0.01 \rho$.

The measurement precision the GW signal amplitude, sky location, and timing
during the eclipse are (e.g., \cite{PoissonWill1995,Krishnan2004,Fairhurst2009,Mandel2017}) 
\begin{eqnarray}
\frac{\delta h}{h} &\sim& \frac{1}{\rho_\mathrm{ec}} \, ;\\
\delta \theta &\sim& \frac{1}{\rho_\mathrm{ec}}
   \frac{c \tau_\mathrm{GW}}{v_\oplus \tau_\mathrm{ec}}
   \simeq \frac{600''}{\rho_{\rm ec}}\left(\frac{1000~\mathrm{Hz}}{f_{\rm GW}}\right)\left(\frac{1~\mathrm{hr}}{\tau_{\rm ec}}\right)\,;\quad \\
\delta t &\sim& \frac{\tau_\mathrm{GW}}{\rho_\mathrm{ec}}=\frac{1~\mathrm{ms}}{\rho_{\rm ec}}\left(\frac{1000~\mathrm{Hz}}{f_{\rm GW}}\right) \, ,
\end{eqnarray}
where $\tau_\mathrm{GW} = 1/f_\mathrm{GW}$ is the GW period, $\tau_\mathrm{ec}$
is the duration of the observation, and $v_\oplus = 2\pi~\mathrm{au\,yr^{-1}}$
is the orbital speed of the Earth.  This allows us to compute the detectability
of these quantities:
\begin{eqnarray}
\frac{\Delta h}{\delta h} &\sim& 0.01 \rho_\mathrm{ec} \, ;\\
\frac{\Delta \theta}{\delta \theta} &\sim&
   0.002\rho_\mathrm{ec}\left(\frac{f_{\rm GW}}{1000~\mathrm{Hz}}\right)\left(\frac{\tau_{\rm ec}}{1~\mathrm{hr}}\right) \, ;\\
\frac{\Delta t}{\delta t} &\sim& 0.01 \rho_\mathrm{ec}\left(\frac{f_{\rm
   GW}}{1000~\mathrm{Hz}}\right) \, .
\end{eqnarray}
As can be seen, for very high-frequency signals with $f_{\rm GW}\sim
10^3~\mathrm{Hz}$ and a time of observation of $\sim 1~\mathrm{hr}$ the
detectability of the three effects are comparable in this order-of-magnitude
analysis.

In order to make a useful measurement, a quantity such as $\Delta
t/\delta t$ should at the least exceed $1$.  This corresponds to the
requirement $\rho_\mathrm{ec} > 100$ for a signal with a GW frequency of $\sim
1000~\mathrm{Hz}$.  For eclipses by the Sun, this means that the full SNR must
be $\rho > 2500$; for eclipses by the AMXP's companion lasting $10\%$ of the
orbit, the requirement is a more modest $\rho>300$.  The SNR is expected to
improve by a factor of $\sim 30$ -- $50$ at high frequencies between the
Advanced LIGO second observing run sensitivity -- the latest data for which
continuous-wave upper limits are available \cite{Abbott+2019blind} -- and
next-generation detectors such as the Einstein Telescope \cite{Hild+2011} and
the Cosmic Explorer \cite{Abbott+2017-CE}.  Moreover, since semi-coherent
searches with segments of length $T_\mathrm{seg}$ between half an hour and of
order of one week were typically used in the past \cite{Abbott+2019blind}, the
optimal coherent SNR would naturally be a factor of
$(\mathrm{yr}/T_\mathrm{seg})^{1/4} \approx 3$ to $10$ times greater.  Thus, if
there are favourably located sources at high frequencies just below the latest
upper limits (corresponding to $\rho \gtrsim 10$, which would be translated to
$\rho \gtrsim 1000$ with an optimal coherent analysis of next-generation data),
it should be possible to observe eclipse signatures with next-generation
detectors.

\section{Conclusions}
\label{conclusions}
We demonstrated how the observation of an eclipsing GW source provides unique
information on the mass distribution of the eclipsing object, and showed that
{CWs with frequencies $>100~\mathrm{Hz}$} are best suited for this purpose.  For
the case of a source eclipsed by the Sun, a GW signal would experience an
apparent change in source position of a few arcseconds, a change in strain
amplitude by up to $\sim 4\%$ and a Shapiro time delay of up to $\sim
0.034~\mathrm{ms}$.  An even more likely possibility is that a source of GWs is
eclipsed by a binary companion, in which case the only signature of the eclipse
for an observer on Earth would be a variable time delay of a magnitude similar
to that produced by the Sun.

The potential to observe lensing of CW sources depends upon the currently
unknown amplitude of their GW emission.  No CW sources have yet been detected.
The effects of eclipses on the signal are small.  Moreover, the SNR accumulated
during the eclipse is a factor of $\sim 3$ to $\sim 25$ smaller than the total
SNR.  Therefore, eclipse lensing may be safely neglected for observations with
current-generation detectors.  However, if CW sources are just at the limit of
the sensitivity of early current-generation detectors
\cite{Abbott_2019_obsruns}, eclipses could potentially be observed by
next-generation detectors.

To obtain interesting constraints on the mass distribution of the eclipsing
object will require SNRs of order of $10^3$.  Such large SNRs may motivate the
development of specialised detectors.  Since a much larger SNR is required to
measure the stellar interior than to detect a signal, we will know ahead of time
the frequency of interest.  This makes eclipsing CW sources an attractive target
for tuneable narrow-band detectors.  These can achieve enhanced sensitivity in a
small range of frequencies compared to wide-band detectors.  Tuneable detectors
have been suggested for observations of binary neutron star coalescences, where
they can track the inspiral and observe the post-merger signal
\cite{Simakov2014,Hughes2002,Graham+2016}, or to increase detection prospects
for supernovae \cite{Srivastava+2019}.  In comparison, eclipsing CW sources are
a far simpler target, since detectors only need to focus on a single known
frequency. Additionally, the timing of eclipses by the Sun can be predicted
years in advance, so tuning does not need to be done dynamically, but can follow
a well-planned schedule.  Observations from multiple eclipsing sources could be
combined to give a more detailed map of the Sun's interior.


While eclipsing CWs provide a new probe of stellar interiors, we have not
addressed the measurement precision necessary to provide meaningful constraints
on stellar structure. For the Sun, helioseismology and neutrino detections
already provide stringent constraints (cf.\ \cite{Christensen-Dalsgaard2018}).
In this context, unless extremely bright high-frequency sources of CWs are
detected, even next-generation detectors might not be sufficient to improve upon
the known constraints on the structure of the Sun. {Still, additional
work is required to properly quantify by how much our lensing predictions are modified
by uncertainties in solar structure.}
For the case when the
eclipsing star is a binary companion to a GW source, neutrinos are undetectable,
and asteroseismic measurements cannot be performed to the same precision as for
the Sun. In this case, the measurement of eclipses of GW sources could provide a
unique view into their stellar interiors.  Despite its complexity, the detection
of eclipses would provide an unbiased measurement of the mass distribution of a
star, independent of uncertainties such as the composition or nuclear reaction
rates.

\acknowledgments

PM thanks the Kavli Institute for theoretical physics of the University of
California Santa Barbara, together with the participants of the ``Astrophysics
from LIGO's First Black Holes'' program for helpful discussion.  KB is grateful
to Mads Sorenson for providing the Python code used to generate the joint
probability distribution of initial binary parameters.  CPLB thanks Nancy
Aggarwal, Denis Martynov, Haixing Miao for useful discussions on future
detectors.  The authors thank Graham Woan, David Keitel and the anonymous referees for careful comments
on the manuscript.  PM acknowledges support from NSF grant AST-1517753 and the
Senior Fellow of the Canadian Institute for Advanced Research (CIFAR) program in
Gravity and Extreme Universe, both granted to Vassiliki Kalogera at Northwestern
University.  CPLB is supported by the CIERA Board of Visitors Research
Professorship, and by the National Science Foundation under Grant No.\ PHY-1912648. {We would also like to extend our thanks to the two anonymous referees, who provided important feedback to this work.}
This document has been assigned LIGO document number LIGO-P1900236.

\appendix

\section{Wave optics calculations}\label{ap:k}

\subsection{The Kirchhoff integral}
\begin{figure}
   \begin{center}
      \includegraphics[width=\columnwidth]{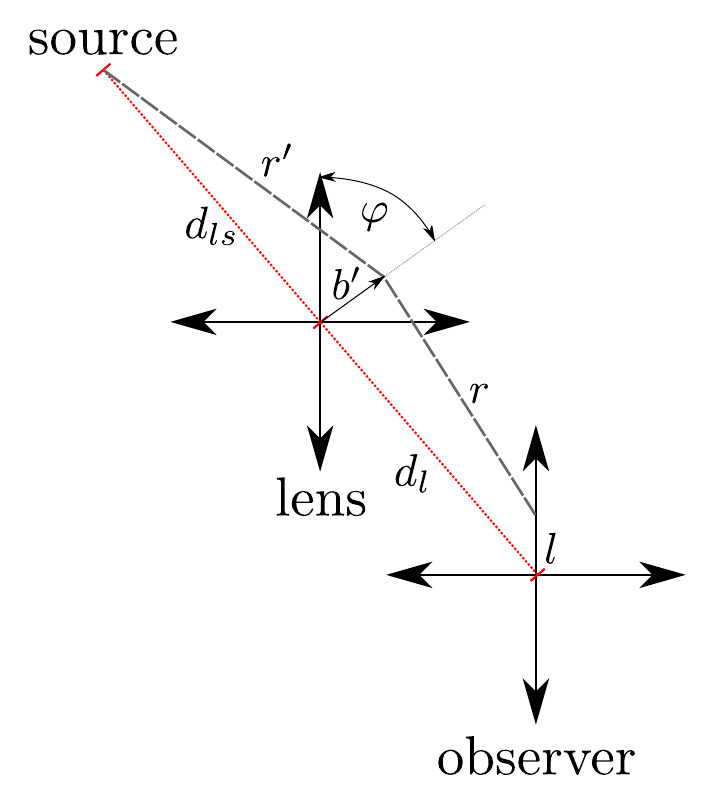}
      \caption{Variables used for the computation of the Kirchhoff integral, in
      order to compute the effect of the Sun on GWs without the
      use of geometric optics. $r$ is the distance
      between the observer and a point in the plane of the lens, while $r'$ is
      the distance between the source and the same point. The distance $l$
      corresponds to the separation between the observer and the optical axis,
      which is the line that joins the source and the lens. The lines
      of length $r$ and $r'$ are not necessarily coplanar, and are not meant to
      represent an actual trajectory (within the geometric optics approximation)
      from the source to the observer.}
      \label{fig:kirchhoff}
   \end{center}
\end{figure}

Following the work of
\cite{Ohanian1974,BontzHaugan1981}, if the amplitude $\Psi$ of a wave is known at a
surface $S$ surrounding the observer and inside which the wave propagates freely, its
amplitude at the observer can be computed using the Kirchhoff integral,
\begin{eqnarray}
   \Psi_o = \frac{1}{4\pi}\int_S\left(\left[\nabla
   \Psi\right]\frac{e^{ikr}}{r}-\Psi\nabla\left[\frac{e^{ikr}}{r}\right]\right)\cdot
   \mathrm{d}\vec{A},
   \label{eq:kirchhoff}
\end{eqnarray}
where $k\equiv 2\pi/\lambda$ is the wavenumber and $r$ is the distance between
the observer and a point at the surface $S$. For the particular case under
consideration, the surface of integration can be taken to be the plane of the
lens (see Fig.~\ref{fig:kirchhoff}). The effect of the lens can be
accounted for as a phase shift due to the time delay for a given impact
parameter $b$,
\begin{eqnarray}
   \Psi(b) = \frac{A_0}{r'}e^{i(kr'-\omega \Delta t(b))},\label{equ:psiplane}
\end{eqnarray}
where $\omega=2\pi f_{\rm GW}$ is the angular frequency of the wave, $r'$ is the
distance between the source and a point in the plane of the lens, and $A_0$ is a
constant that sets the intensity of the wave. To compute the Kirchhoff integral we only consider the case of sources eclipsed by the sun and observed from Earth, in which case $b'$ can be approximated as $b$ (see Fig.~\ref{fig:diagramlens}). Here we ignore the phase shift
$\omega t$ for the time evolution of the wave, as this factors
out in the calculation of time delays and amplitude changes. Assuming $k\ll
d_l^{-1}$ and $k\ll d_{ls}^{-1}$, combining Eq.~\eqref{eq:kirchhoff} and Eq.~\eqref{equ:psiplane} results in
\begin{eqnarray}
      \Psi_o(l) &=& -\frac{ikA_0}{4\pi}\int_0^\infty\int_0^{2\pi}
      e^{i(kr+kr'-\omega\Delta t(b))} \qquad\qquad \nonumber \\
      && \times \frac{1}{rr'}\left(\frac{d_l}{r}+\frac{d_{ls}}{r'}\right) b \,\mathrm{d}\varphi \,\mathrm{d} b,
      \label{equ:psi-limit}
\end{eqnarray}
where
\begin{eqnarray}
      r&=&\sqrt{d_l^2+(l-b\cos\varphi)^2+b^2\sin^2\varphi}, \label{equ:waaaa} \\
      r'&=&\sqrt{d_{ls}^2+b^2},
\end{eqnarray}
$l$ is the distance of the observer from the optical axis while $\varphi$ is
an angle in the optical plane (see Fig.~\ref{fig:kirchhoff}). 
Given $\Psi_o(l)$, the amplification of a GW {arriving} at Earth is
given by the ratio of $\Psi_o(l)$ to the expected amplitude in the absence of a
lens,
\begin{eqnarray}
   \frac{h'(b)}{h}=\frac{|\Psi_o|(d_l+d_{ls})}{A_0},
\end{eqnarray}
while the time delay can be computed by comparing the phase of the wave at
different values of $l$,
\begin{eqnarray}
      \omega|\Delta t_s| = \arccos\left[\frac{\Psi_o(b)\cdot \Psi_o(R_\odot)}{|\Psi_o(b)||\Psi_o(R_\odot)|}\right],
\end{eqnarray}
where $\Psi_o(b)\cdot \Psi_o(R_\odot)$ denotes the complex dot product of $\Psi_o(b)$ and $\Psi_o(R_\odot)$.

The computation of Eq.~\eqref{equ:waaaa} can be simplified in the limit
$l\ll d_l$, in which case
\begin{eqnarray}
      r&=&r_0\left[1+\frac{bl\cos
   \varphi}{r_0^2}-\mathcal{O}\left(\frac{b^2l^2}{r_0^4}\right)\right],\label{equ:approxr0}
\end{eqnarray}
where $r_0=\sqrt{d_l^2+b^2+l^2}$. For the terms outside the exponential factor
in Eq.~\eqref{equ:psi-limit} one can simply approximate $r\simeq r_0$,
but the term of order $\mathcal{O}(bl/r_0^2)$ in Equation (\ref{equ:approxr0})
needs to be included in the exponential to prevent errors in the phase
larger than those induced by the lens. Under these approximations one has that
\begin{eqnarray}
      \Psi_o(l) &=& -\frac{ikA_0}{4\pi}\int_0^\infty\int_0^{2\pi}
      e^{i(kr_0+kr'+kbl\cos\varphi/r_0-\omega\Delta t(b))} \nonumber \\
      && \times \frac{1}{r_0r'}\left(\frac{d_l}{r_0}+\frac{d_{ls}}{r'}\right) b \,\mathrm{d}\varphi \,\mathrm{d} b,
\end{eqnarray}
and the   integral over $\phi$ can be computed analytically resulting in
\begin{eqnarray}
      \Psi_o(l) &=& -\frac{ikA_0}{2}\int_0^\infty
      e^{i(kr_0+kr'-\omega\Delta t(b))}\qquad\qquad \nonumber \\
      && \times
      \frac{J_0(kbl/r_0)}{r_0r'}\left(\frac{d_l}{r_0}+\frac{d_{ls}}{r'}\right) b \,\mathrm{d}
      b,
      \label{equ:waaa2}
\end{eqnarray}
where $J_0$ is the Bessel function of the first kind and order zero.

Previous work on the impact of the Sun on GWs
\cite{Ohanian1974,BontzHaugan1981} dealt with the change in amplitude of the
wave near the caustics of the solar lens (regions where multiple images are formed). In this case the integral in
Eq.~\eqref{equ:waaa2} is dominated by points around the value of $b$ for which geometrical optics predicts rays converge at a distance $d_l$ from the optical plane. This allows an analytical approximation of the result
using the stationary phase approximation \cite{BornWolf1999} and leads to the
conclusion of \cite{Ohanian1974} that no significant amplification occurs 
for frequencies $f_{\rm GW}<10^4$ Hz. The situation is different for an observer at
Earth as the Earth is not located at a caustic of the Sun, requiring the calculation of
the Kirchhoff integral in the entire lens plane; from our solar model we
predict caustics occur at a distance of $\gtrsim 20~\mathrm{au}$ from the Sun.

\subsection{Numerical integration of the Kirchhoff integral}

Numerically computing the integral in Eq.~\eqref{equ:waaa2} is difficult,
as both the exponential term and the Bessel function change sign leading to a
rapidly oscillating integrand. For $b\gg d_0$ we have that $r_0\sim b$, such
that the argument of the Bessel function becomes constant and only the rapid
oscillation of the exponential factor remains problematic. To remedy this, we
make the following change of variables:
\begin{eqnarray}
   y &=& r_0+r'-d_{ls}-\sqrt{d_l^2+l^2},\\
   \mathrm{d} y&=&\left(\frac{1}{r_0}+\frac{1}{r'}\right)b\,\mathrm{d}b,
\end{eqnarray}
which, except for the time delay produced by the lens,
leaves the argument of the complex exponential in Eq.~\eqref{equ:waaa2} 
as the integration variable. Ignoring constant phase shifts
the wave amplitude is then
\begin{eqnarray}
      \Psi_o(l) &=& -\frac{kA_0e^{ik\sqrt{d_l^2+l^2}}}{2}\int_0^\infty
      e^{i(ky-\omega\Delta t(b))}\qquad\qquad \nonumber \\
      && \times
      \frac{J_0(kbl/r_0)}{r_0+r'}\left(\frac{d_l}{r_0}+\frac{d_{ls}}{r'}\right) 
      \mathrm{d} y.
\end{eqnarray}
We can then perform the integrand over each individual cycle produced by the
term $ky$ in the complex exponential, defining for an integer value $j$ the
quantity
\begin{eqnarray}
      A_j &=& -\frac{kA_0}{2}\int_{2\pi j/k}^{2\pi(j+1)/k}
      e^{i(ky-\omega\Delta t(b))}\qquad\qquad \nonumber \\
      && \times
      \frac{J_0(kbl/r_0)}{r_0+r'}\left(\frac{d_l}{r_0}+\frac{d_{ls}}{r'}\right) 
      \mathrm{d} y,\label{equ:Aj}
\end{eqnarray}
such that
\begin{eqnarray}
   \Psi_o(l) = e^{ik\sqrt{d_l^2+l^2}}\sum_{j=0}^{\infty}A_j.
\end{eqnarray}
If there exists an $N$ such that the value of $A_j$ changes slowly with $j$ for $j>N$, then
it is useful to separate the sum to include all terms up to $N$, and express the
rest as an integral,
\begin{eqnarray}
   \sum_{j=0}^{\infty}A_j &=& \sum_{j=0}^{N}A_j + \int_{N+1}^{\infty} A(x)\,\mathrm{d}x
\end{eqnarray}
where $A(x)$ is a function that is equal to $A_j$ with $j={\rm floor}(x)$
being the nearest integer to $x$ that is smaller than $x$. Switching variables to $z=\ln x$ in
the integral results in
\begin{eqnarray}
   \sum_{j=0}^{\infty}A_j&=&\sum_{j=0}^{N}A_j + \int_{\ln(N+1)}^{\infty}\frac{A(e^z)}{e^z}\,\mathrm{d}z.
\end{eqnarray}
Since $A_j$ varies slowly for $j>N$, the integral can be estimated by adding
over logarithmic intervals $\Delta z$,
\begin{eqnarray}
   \sum_{j=0}^{\infty}A_j&\simeq&\sum_{j=0}^{N}A_j +
   \sum_{m=0}^{\infty}\frac{A(e^{\ln(N+1)+m\Delta z})}{e^{\ln(N+1)+m\Delta z}}\Delta z.
\end{eqnarray}
This allows the calculation of the integral up to a large number of cycles,
without individually computing the contribution of each one. Using this, we
numerically compute the real and imaginary part of $\Psi_0(l)$. In addition,
when evaluating cosines or sines in Eq.~\eqref{equ:Aj}, rather than computing,
for example, $\cos(ky-\omega \Delta t)$ we compute instead $\cos[k(y-2\pi
j/k)-\omega \Delta t]$. This prevents the evaluation of trigonometric functions
with large arguments and reduces numerical errors.

\begin{figure}[ht!]
   \begin{center}
      \includegraphics[width=\columnwidth]{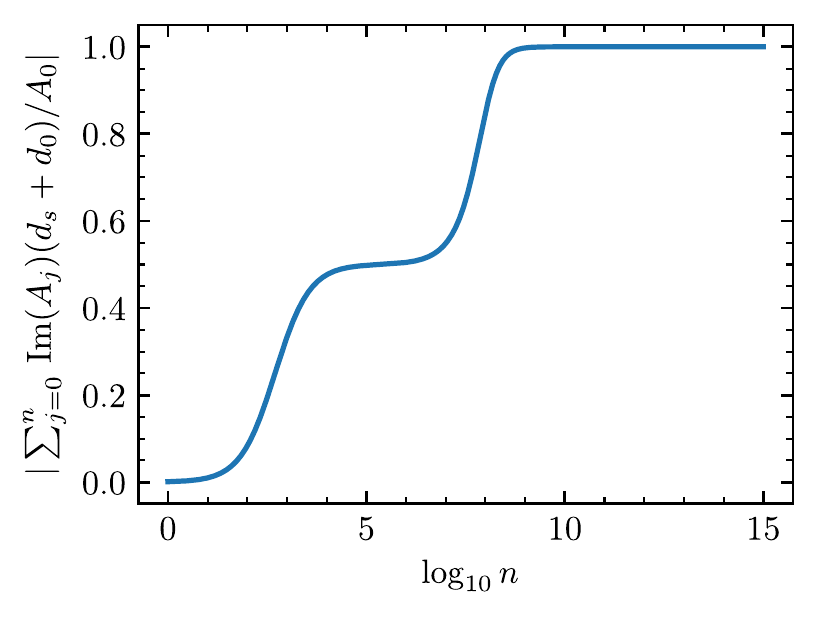}
      \caption{Computation of the imaginary part of $\sum A_j$ for
      $l=0.05R_\odot$, considering a $100$ Hz source located at $1$ parsec and observed
      at $1$ au from the Sun. The ordinate-axis is normalized to the expected intensity
      of the wave in the absence of a lens. The abscissa-axis
      represents the number of cycles added (see Appendix A).}
      \label{fig:integral_hell}
   \end{center}
\end{figure}

Figure~\ref{fig:integral_hell} shows an example of this integration, showing only
the imaginary part of $\Psi_o$ as the number of cycles $n$ included in the
calculation of the integral is increased. In this particular example the
integral is directly computed up to $N=10^6$, and then estimated up to $10^{15}$
cycles using $10^6$ equally spaced logarithmic intervals.  As it can be seen, the integral
converges after $10^{10}$ cycles. Directly computing the integral up to that
point is extremely expensive, which is the reason why we require the
approximation discussed in this Appendix. Evaluating Eq.~\eqref{equ:Aj} still requires a choice for $d_{ls}$ and $d_l$. The calculations shown in Section~\ref{sec:wave} were done using a distance between the lens and the source of $d_{ls}=1\;\mathrm{pc}$, but we have verified that the resulting amplification and time delays are equivalent if the integrations are done using $d_{ls}=0.1\;\mathrm{pc}$.

%


\end{document}